\documentclass[aps,pra,amsmath,reprint,floatfix,groupedaddress]{revtex4-1}

\usepackage{siunitx}
\newcommand{\equationref}[1]{Eq.~(\ref{#1})}

\usepackage{graphicx}
\usepackage{amsmath}
\usepackage{float}

\begin{document}

\title{Precision measurement of the ionization energy and quantum defects of $^{39}$K \small{I}}


\author{Michael Peper}
\email{michael.peper@phys.chem.ethz.ch}
\author{Felix Helmrich}
\author{Jonas Butscher}
\author{Josef Anton Agner}
\author{Hansj\"urg Schmutz}
\author{Fr\'ed\'eric Merkt}
\email{merkt@phys.chem.ethz.ch}
\author{Johannes Deiglmayr}
\altaffiliation{Present address: University of Leipzig, Felix-Bloch Institut, Leipzig, Germany}
\email{johannes.deiglmayr@physik.uni-leipzig.de}
\affiliation{Laboratory of Physical Chemistry, ETH Z\"urich, 8093 Z\"urich, Switzerland
}


\date{\today}

\begin{abstract}
We present absolute-frequency measurements in ultracold $^{39}$K samples of the transitions from the $4s_{1/2}$ ground state to $np_{1/2}$ and $np_{3/2}$ Rydberg states. A global nonlinear regression of the $np_{1/2}$ and $np_{3/2}$ term values yields an improved wave number of \num{35009.8139710(22)}$_\mathrm{sys}$(3)$_\mathrm{stat}$\,\si{\per\centi\meter} for the first ionization threshold of $^{39}$K and the quantum defects of the $np_{1/2}$ and $np_{3/2}$ series. In addition, we report the frequencies of selected one-photon transitions $n's_{1/2}\leftarrow np_{3/2}$, $n'd_{j}\leftarrow n p_{3/2}$, $n'f_{j'}\leftarrow n d_{j}$ and $n'g_{j'}\leftarrow n f_{j}$ and two-photon transitions $nf_{j'}\leftarrow  n p_{j}$ determined by millimeter-wave spectroscopy, where $j$ is the total angular momentum quantum number. By combining the results from the laser and millimeter-wave spectroscopic experiments, we obtain improved values for the quantum defects of the $s_{1/2}$, $d_{3/2}$, $d_{5/2}$, $f_{j}$ and $g_{j}$ states. For the $d_j$ series, the inverted fine structure was confirmed for $n \geq 32$. The fine-structure splitting of the $f$ series is less than \SI{100}{kHz} at $n=31$, significantly smaller than the hydrogenic splitting, and the fine structure of the $g$ series is regular for $n \geq 30$, with a fine-structure splitting compatible with the hydrogenic prediction. From the measured quantum defects of the $f$ and $g$ series we derive an estimate for the static dipole $\alpha_\mathrm{d}$ and quadrupole $\alpha_\mathrm{q}$ polarizabilities of the K$^+$ ion core. Additionally, the hyperfine splitting of the $4s_{1/2}$ ground state of $^{39}$K was determined to be \SI{461.719700(5)}{MHz} using radio-frequency spectroscopy and Ramsey-type interferometry.
\end{abstract}

\pacs{}

\maketitle

\section{Introduction}
High-resolution spectroscopy of atomic and molecular systems provides important information about atomic and molecular energy levels, from which physical-chemical properties such as the core polarizability \cite{freemankleppnerpra1976}, isotope shifts \cite{aldridgepra2011} or molecular dissociation energies \cite{beyerpra2018} may be derived. Precise measurements of the binding energies of the Rydberg states of atoms and molecules enable the accurate determination of long-range interaction potentials between Rydberg atoms~\cite{marcassaChapterTwoInteractions2014,deiglmayrLongrangeInteractionsRydberg2016,weberCalculationRydbergInteraction2017,arc2017} and ionization energies by Rydberg-series extrapolation \cite{deiglmayrpra2016} and multi-channel quantum defect theory \cite{sprecherMeasuringIonisationDissociation2011}.

The line spectra of potassium were observed already in the early years of atomic spectroscopy, as summarized in the work of Fowler \cite{fowler1922} and Paschen and G\"otze \cite{paschen1922seriengesetze}. The work of \'Edlen \cite{ Edlenzeitphys1936}, Kratz \cite{Kratzphysrev1949} and Risberg \cite{risbergarkfys1956} was later improved and extended by Lorenzen \textit{et al.} for the $ns_{1/2}$, $np_{j}$ and $nd_j$ series \cite{LORENZENoptcomm1981,Lorenzenphysscrip1983}. Quantum defects, polarizabilities and fine-structure intervals of the $nd$ Rydberg series of potassium were studied by high-resolution millimeter-wave spectroscopy \cite{gallagherpra1978}. Among the alkali atoms, potassium is the only element with three naturally abundant isotopes: $^{39}$K, $^{40}$K, and $^{41}$K. Whereas $^{39}$K and $^{41}$K are bosons, $^{40}$K is a metastable fermion. The isotopic shifts of the lowest states of potassium have been investigated by \citet{pendrillIsotopeShiftsEnergy1982}.

In this article, we present an experimental study of the $ns$, $np$, $nd$, $nf$, and $ng$ Rydberg series of $^{39}$K, which combines laser-cooled samples of ultracold atoms, frequency-comb-referenced UV spectroscopy, and millimeter-wave spectroscopy. The article is structured as follows: after introducing the experimental setup in Section~\ref{Experiment}, the origins of systematic uncertainties in our setup are analyzed (Section~\ref{systematicshifts}). The UV-spectroscopy of $np$ Rydberg states, the extraction of quantum defects and the determination of the ionization energy are discussed in Section~\ref{optspectroscopyp}. The results of the millimeter-wave study of the $ns$, $nd$, $nf$, and $ng$ series are presented in Section~\ref{mmwavespec}. In the course of this study, we also remeasured the hyperfine splitting of the 4$s_{1/2}$ ground-state of $^{39}$K (Section~\ref{gssplitting}) and derived values for the static dipole and quadrupole polarizabilites of $^{39}$K$^+$ from the quantum defects of the $f$ and $g$ series (Section~\ref{corepolarization}).

\section{Experiment}
\label{Experiment}
For the work presented in this article, the existing setup for high-resolution spectroscopy of cesium Rydberg atoms~\cite{sassmannshausen13,deiglmayrpra2016} was extended and modified to allow also for the high-resolution spectroscopy of potassium atoms. The atomic-vapor source was replaced by a double-species oven~\cite{stanMultipleSpeciesAtom2005}, loaded from ampoules containing pure alkali metals, which were cracked under vacuum by squeezing thin copper tubes enclosing the ampoules. The temperatures of the cesium reservoir, the potassium reservoir, and the mixing zone are typically kept at \SI{24}{\celsius}, \SI{54}{\celsius}, and \SI{60}{\celsius}, respectively. The oven is connected to the main chamber by an all-metal corner valve, which is used to adjust the background vapor pressure in the main chamber to approximately \SI{1e-10}{\milli\bar}.

Light for laser-cooling of potassium is provided by an amplified diode laser system (\textsc{Toptica} DLC TA Pro 767, output power \SI{2}{\watt}), stabilized by Doppler-free saturated-absorption spectroscopy of potassium vapor in a quartz cell. The laser beams with different frequencies for cooling, repumping, and absorption imaging are derived from the master laser using acousto-optic modulators, which are also used for modulating the intensity of the laser beams. The laser beams creating the magneto-optical traps (MOTs) for cesium and potassium are overlapped on dichroic beam splitters and are coupled into three separate polarization-maintaining fibers for the three spatial axes. In front of the chamber, the light is coupled out of the fibers, collimated to a waist radius of about \SI{1}{\cm}, and converted into circulary-polarized light by achromatic $\lambda/4$-wave plates. Sub-Doppler cooling of potassium is achieved by a dark-optical-molasses scheme~\cite{gokhrooSubDopplerDeepcooledBosonic2011,landiniSubDopplerLaserCooling2011}.

 The resulting samples typically contain \num{e7} $^{39}$K atoms at densities of about \SI{2e10}{\per\cubic\centi\metre} and have a translational temperature of \SI{20}{\micro\kelvin}. Magnetic offset fields are controlled by applying currents to three external pairs of coils in Helmholtz configuration \cite{sassmannshausen13}. Radio-frequency spectroscopy of the hyperfine interval ($F=2\leftarrow F=1$) in the $4s_{1/2}$ ground-state of $^{39}\mathrm{K}$ is employed to determine the coil currents which minimize the quadratic Zeeman shift of the $F=2,m_F=0 \leftarrow F=1,m_F=0$ transition and thus the magnitude of the magnetic field experienced by the atoms. The residual field is below \SI{7}{mG}, as estimated from the \SI{1}{mA} accuracy of the current source used for magnetic-field compensation. 
 
The atoms are prepared in either the lower ($F=1$) or the upper ($F=2$) hyperfine component of the $4s_{1/2}$ ground state. Subsequently, they are excited to $np_{j}$ Rydberg states by pulses of frequency-tunable light at 285--\SI{288}{\nano\metre}. The light is obtained by intracavity frequency doubling (\textsc{Coherent MBD 200}) the output of a ring dye laser (\textsc{Coherent 899-21} operated with the dye Rhodamine~6G), which is pumped by a frequency-doubled continuous-wave Nd:YVO$_4$ laser (\textsc{Laser Quantum finesse 532}), and shaped into pulses using an acousto-optic modulator. Electric fields for stray-field compensation and pulsed-field ionization (PFI) are applied to the Rydberg atoms in the two different configurations described by \citet{sassmannshausen13}. In configuration I, a pair of segmented ring-shaped electrodes is used for efficient field compensation in all three spatial directions. Electric stray fields are compensated frequently to below \SI{6}{mV/cm} by a measurement of the quadratic Stark effect of the $130p_{3/2}$ state. An additional electrode in front of the ion detector (see below) is used to create the large electric-field pulses required for state-selective PFI. In this configuration, the maximal electric field only efficiently ionizes Rydberg states with $n \gtrsim 50$. In configuration II, we apply the electric-field pulses for PFI to four of the eight segments of the ring-shaped electrodes used for field compensation. This allows the field ionization of Rydberg states with $n$ as small as 30 at the cost of a less complete compensation of stray electric fields \cite{sassmannshausen13}. States below $n\approx30$ can still be observed by detecting spontaneously formed K$^+$ ions \cite{sassmannshausen13}.

For a first set of measurements (labelled I below) the frequency of the ring dye laser is locked to an external reference cavity (\textsc{Thorlabs} SA200-5B). The absolute frequency of the UV light is obtained by measuring the frequency of the fundamental laser light with a frequency comb (\textsc{Menlo Systems FC1500-250-WG}) \cite{deiglmayrpra2016}. In this measurement series, the electric-field-compensation configuration I was employed. In a second set of measurements  (labelled II below), the ring-dye-laser frequency is directly locked to the frequency comb \cite{beyerpra2018}. In the measurements series II, and for all measurements on \textit{s, d, f} and \textit{g} Rydberg states by millimeter-wave spectroscopy, the electric-field-compensation configuration II was used.

The K$^+$ ions resulting from PFI are accelerated towards, and detected on, a microchannel-plate detector \cite{sassmannshausen13}. Transitions to $60p_{3/2}$ and $70p_{3/2}$ Rydberg states have been measured with both configurations I and II to verify the consistency of both sets of measurements. The transition to the $60p_{3/2}$ state is recorded regularly to detect time-dependent systematic shifts. The measured transition frequency varies over time with an amplitude of about \SI{50}{kHz}, significantly larger than the typical statistical uncertainty of \SI{20}{kHz}. This fluctuation can be traced back to a systematic error in the frequency calibration in measurement configuration I caused by a ground loop in the locking electronics. This systematic calibration error is removed in configuration II.

For the determination of quantum defects of the $s$, $d$, $f$ and $g$ Rydberg states, transitions between Rydberg states are recorded by millimeter-wave spectroscopy. The general procedure is described in Refs.~\cite{merkt1998very,sassmannshausen13,deiglmayrpra2016}. In the present work, the outputs of three radio-frequency (RF) generators (\textsc{Wiltron 6769B},  \textsc{Agilent E8257D} and \textsc{Anritsu MG3692A}) are either used in the fundamental or after harmonic generation using an active 6-fold  (70--\SI{110}{GHz}, \textsc{OML Inc. S10MS-AG}), 12-fold (110--\SI{170}{GHz}, \textsc{Virginia Diodes WR-6.5}) or 18-fold (170--\SI{250}{GHz}, \textsc{Virginia Diodes WR-9.0} with \textsc{Virginia Diodes WR4.3x2}) multiplier. Pulses are formed by modulating the fundamental output of the RF generators. The millimeter-wave radiation is coupled to free space by suitable horns and sent into the vacuum chamber through an optical viewport. The power of the millimeter-wave radiation is adjusted by inserting calibrated, adjustable attenuators after the harmonic-generation units, and adding additionally stacks of paper in front of the chamber. The induced population transfer between the Rydberg states is detected by state-selective PFI.

The hyperfine splitting ($F=2 - F=1$) of the $4s_{1/2}$ state was also determined by RF spectroscopy on atoms released from the magneto-optical trap. For this measurement, the pulsed output of a RF generator (\textsc{HP 8647A}) is amplified to \SI{30}{dBm} and applied to one of the electrodes used for electric-field compensation. The population in the $F=2$ level is probed by absorption imaging. All RF generators, as well as the optical frequency comb, are referenced to a GPS-disciplined Rubidium atomic clock (\textsc{Stanford Research Systems FS725} with a \textsc{Spectrum Instruments TM-4} GPS receiver).

\section{Systematic frequency shifts}
\label{systematicshifts}
In contrast to our previous measurements on Cs \cite{deiglmayrpra2016}, the samples with which the measurements are performed in this work are not trapped in an optical dipole trap (ODT). Hence systematic frequency shifts resulting from the AC Stark shift caused by the ODT trapping laser, which were an important source of systematic uncertainty in Ref.~\cite{deiglmayrpra2016}, are absent. Table~\ref{errors} summarizes the main sources of uncertainties of our measurements.

The power of the UV-laser pulses was chosen such that no systematic dependence of the transition frequencies on the excitation intensity was observed. This measure excludes significant ($>\SI{10}{kHz}$) contributions from the AC Stark shift caused by the excitation laser, Rydberg-Rydberg interactions, and Rydberg-ion interactions to the observed transition frequencies. The AC Stark shift caused by thermal radiation from the environment at room temperature (measured to be \SI{2.4}{\kilo\hertz} at \SI{300}{\kelvin} \cite{hollbergprl1984}) is also negligible at the precision level of our experiment.

As experimentally observed by Amaldi and Segr\'e \cite{Amaldiilnuovocimento1933} and \citet{fuchtbauer1923intensitat}, a Rydberg atom within a gas of ground-state atoms experiences a shift dependent on the gas pressure. This shift was interpreted by Fermi \cite{fermiilnuovo1934} as arising from low-energy scattering of the Rydberg electron off ground-state atoms located within its orbit. Using the triplet $s$-wave scattering length for e$^-$-K collisions of \SI{-15.4}{\textit{a}_0} \cite{fabrikantjphysb1986} and a peak density of \SI{2E10}{\per\cubic\centi\meter}, we obtain a maximum shift of \SI{2}{\kilo\hertz} at high values of $n$.

Prior to laser excitation of the $np$ Rydberg states, the atomic cloud is released from the MOT and accelerated in vertical direction by gravitational forces. Because the UV laser also propagates in vertical direction, this causes a first-order Doppler shift. Our setup images the atomic density only in the horizontal plane and is thus blind to translations in the vertical direction. In the final steps of the sample preparation, magnetic compression and molasses cooling is applied, making the exact velocity distribution of the atoms at the time of UV excitation difficult to predict. We estimate an upper bound for the Doppler shift of \SI{35}{kHz}, based on the delay between the beginning of molasses cooling and the UV-excitation pulse. Higher-order Doppler shifts are negligible.

The photon-recoil shift \cite{Hall1976,Kolchenko1969} is an important contribution to the observed transition frequencies. This shift results from the energy and momentum conservation in the absorption process and is given by
\begin{equation}
\label{eq:recoilshift}
    \Delta\nu=\frac{\Delta E(f\leftarrow i)}{h}-\nu=-\frac{h\nu^2}{2mc^2},
\end{equation}
where $\Delta E(f\leftarrow i)$ is the energy seperation between the initial and final states of the absorbing species, $m$ its mass, $\nu$ is the frequency of the absorbed photon, $h$ is Planck's constant, and $c$ is the speed of light. For $np_j\leftarrow4s_{1/2}$ transitions into high-lying Rydberg states in $^{39}$K, this shift is roughly \SI{60}{\kilo\hertz}. Because the contribution from the photon-recoil shift can be calculated exactly, this shift does not increase the uncertainty of our measurements. For the millimeter-wave transitions $n'l'_{j'}\leftarrow nl_j$, and the transition between the two ground-state hyperfine components, the photon recoil is negligible. All transition energies reported in this article have been corrected for the photon-recoil shift.

As discussed in Section~\ref{Experiment}, frequencies determined in measurement series I suffer from a systematic calibration error. By intentionally adding a corresponding offset to the frequencies from measurement series I in the global analysis, we estimate the contribution of this calibration error to the uncertainty of the ionization energy to be \SI{50}{\kilo\hertz}, which dominates the overall uncertainty.

\begin{table}
\caption{Systematic errors in the determination of the ionization energy of $^{39}$K \small{I}.}
\label{errors}
\begin{ruledtabular}
\begin{tabular}{lc}
&  $\left|\Delta\nu\right|$ (kHz) \\
\hline
DC Stark shift (negative) &  $<27$  \\
Excitation-power dependent shifts &  $<10$ \\
Zeeman shift &  $<2$ \\
Pressure shift (negative) &  $<2$\\
1$^\mathrm{st}$ order Doppler shift (positive) &  $<35$ \\
2$^\mathrm{nd}$ order Doppler shift &  $<1$ \\
Frequency-calibration error &  50 \\
\hline
Total systematic uncertainty & 67
\end{tabular}
\end{ruledtabular}
\end{table}

\section{Ionization energy and $p$-series quantum defects}
\label{optspectroscopyp}

The extended Ritz formula gives an accurate description of the term values of an unperturbed Rydberg series
\begin{equation}
\tilde{\nu}_{nlj}=\frac{1}{hc}E_\mathrm{I}-\frac{R_\mathrm{K}}{n^{*2}}=\frac{1}{hc}E_\mathrm{I}-\frac{R_\mathrm{K}}{\left[ n-\delta_{lj}(n)\right]^{2}},\label{eq:rydberg}
\end{equation}
with energy-dependent quantum defects
\begin{widetext}
\begin{equation}
\delta_{lj}(n)=\delta_{0,lj}+\frac{\delta_{2,lj}}{\left[ n-\delta_{lj}(n)\right]^2}+\frac{\delta_{4,lj}}{\left[ n-\delta_{lj}(n)\right]^4}+\frac{\delta_{6,lj}}{\left[ n-\delta_{lj}(n)\right]^6}+\dots\;.\label{eq:seriesexpansion}
\end{equation}
\end{widetext}
$E_\mathrm{I}$ is the first ionization energy, $R_\mathrm{K}$ is the reduced Rydberg constant, $n$ is the principal quantum number, $l$ is the orbital angular-momentum quantum number and $j$ is the total angular-momentum quantum number. For $^{39}$K, $R_\mathrm{K}=\SI{109735.7706656(7)}{cm^{-1}}$ using the values for the natural constants from Ref.~\cite{nistcodata} and the mass of $^{39}$K \cite{nistmasses}. Because \equationref{eq:seriesexpansion} is defined recursively (the right-hand side depends on the quantum defect $\delta_{lj}(n)$), it cannot be used directly in a nonlinear-regression analysis. The series expansion is thus modified as described by Drake and Swainson \cite{drakeswainsonpra1991} by replacing $\delta_{lj}(n)$ on the right-hand side by the zeroth-order term $\delta_{0,lj}$, yielding
\begin{widetext}
\begin{equation}
\delta_{lj}(n)=\delta_{0,lj}+\frac{\delta_{2,lj}}{\left[ n-\delta_{0,lj}\right]^2}+\frac{\delta_{4,lj}}{\left[ n-\delta_{0,lj}\right]^4}+\frac{\delta_{6,lj}}{\left[ n-\delta_{0,lj}\right]^6}+\dots\label{eq:seriesexpansionmod} \;.
\end{equation}
\end{widetext}

\begin{figure}
	\includegraphics[trim={0 0cm 0 0},clip,width=\linewidth]{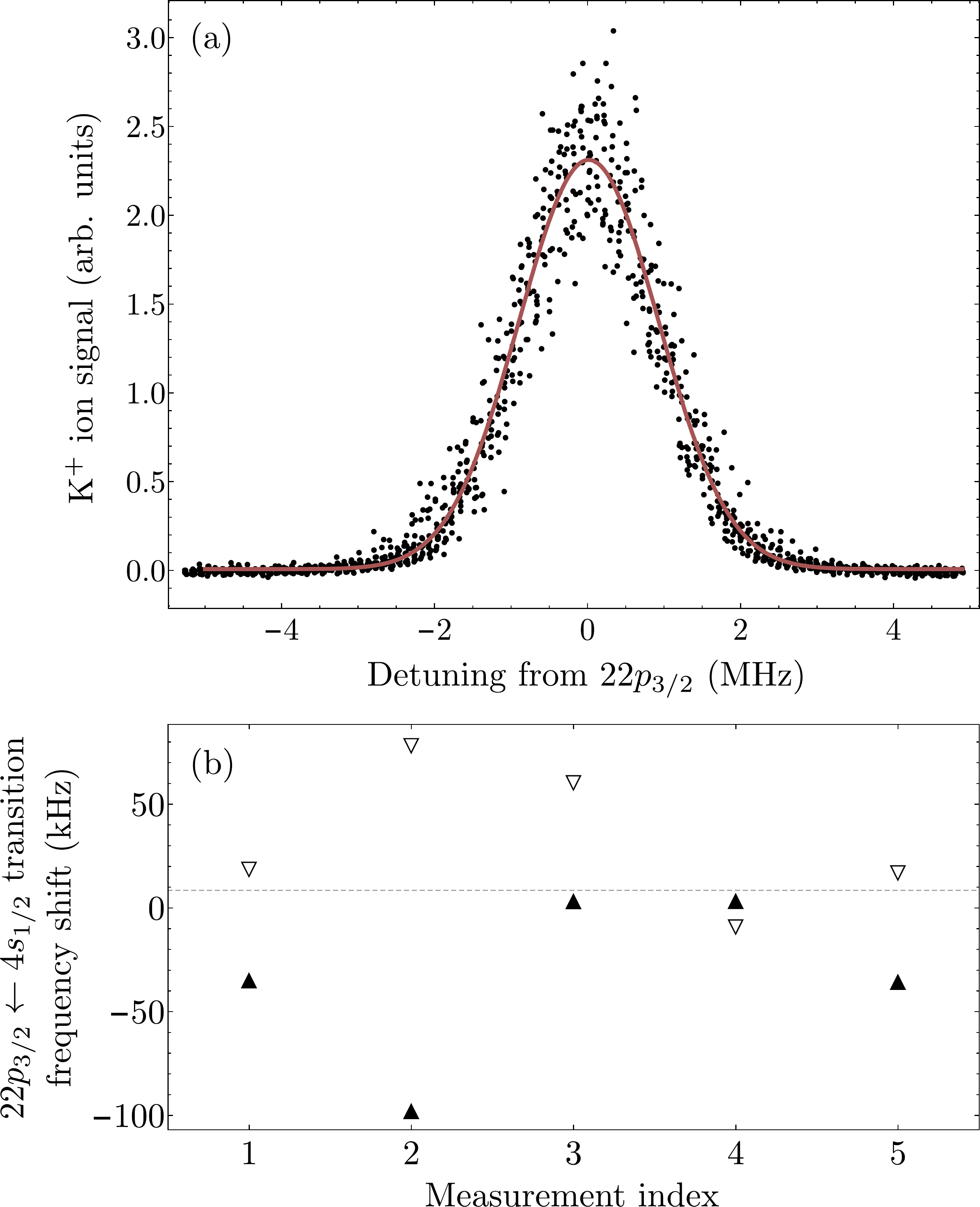}
	\caption{\label{fig:56ptrans} (a) K$^+$ ion signal as a function of the detuning from the $22p_{3/2}\leftarrow4s_{1/2}(F=1)$ transition frequency in MHz (black dots). The red line represents a Gaussian line fit to the observed data points shown in black. (b) Transition frequencies of the five individual measurements from low to high (filled triangles) and high to low (open triangles) frequencies. Zero detuning corresponds to the predicted transition frequency to the $22p_{3/2}$ state using Eqs.~(\ref{eq:rydberg}) and (\ref{eq:seriesexpansionmod}) with the expansion coefficients given in Table~\ref{qdp}. The grey horizontal dashed line indicates the transition frequency determined from the fit of a Gaussian to the full set of data points.}
\end{figure}

\begin{table*}
	\caption{Wave numbers and fit residuals of all $np_{j}\leftarrow4s_{1/2}$ transitions included in the determination of the ionization energy and the $p_j$ quantum defects of $^{39}$K. The transitions are given relative to the center of gravity of the $4s_{1/2}$ ground-state level and are corrected by the photon-recoil shift. The labels I and II refer to the measurement configurations discussed in Section~\ref{Experiment}. Additional transitions from Refs.~\cite{falkepra2006,Johansson1972,Lorenzenphysscrip1983,risbergarkfys1956} are included in the determination of the ionization energy and quantum defects of the $np_j$ series. Our uncertainties correspond to one standard deviation (see text). \label{ptrans}}
			\begingroup\renewcommand{\arraystretch}{1}
	\begin{tabular}{c S[table-format=6.10] S[table-format=8.0] S[table-format=6.10] S[table-format=8.2] l}
		\hline\hline
		\multicolumn{1}{c}{$\vphantom{\frac{\frac{a}{b}}{\frac{a}{b}}} n$}& \multicolumn{1}{c}{$\tilde{\nu}_{p_{1/2}}$ (cm$^{-1}$) }&  \multicolumn{1}{c}{$\delta_\mathrm{fit}$ (kHz)} & \multicolumn{1}{c}{$\tilde{\nu}_{p_{3/2}}$ (cm$^{-1}$)} & \multicolumn{1}{c}{$\delta_\mathrm{fit}$ (kHz)} & Ref.\\
		\hline
4   & 12985.1851949(21)     & 0       & 13042.8954964(39)     & 0   &   \cite{falkepra2006} \\
5   & 24701.382(5)  & 3232    & 24720.139(5)  & -317   & \cite{Johansson1972} \\
6   & 28999.27(3) & -758488 & 29007.71(3) & -396603 &\cite{risbergarkfys1956}\\
7   & 31069.90(3) & -223561 & 31074.40(3) & -101983 &\cite{risbergarkfys1956}\\
8   & 32227.44(3) & 94949   & 32230.11(3) & -101881 & \cite{risbergarkfys1956}\\
9   & 32940.2030(30)  & 17253   & 32941.9262(30)  & 95916  &\cite{Lorenzenphysscrip1983} \\
10  & 33410.2306(30)  & 128962  & 33411.3986(30)  & 46696  &\cite{Lorenzenphysscrip1983} \\
11  & 33736.4979(30)  & 88339   & 33737.3284(30)  & 33651  & \cite{Lorenzenphysscrip1983}\\
12  & 33972.2064(30)  & 129707  & 33972.8148(30)  & 153    & \cite{Lorenzenphysscrip1983}\\
13  & 34148.0284(30)  & 228296  & 34148.4861(30)  & 38005  & \cite{Lorenzenphysscrip1983}\\
14  & 34282.6573(30)  & 143740  & 34283.0181(30)  & 172829 &\cite{Lorenzenphysscrip1983} \\
15  & 34388.0315(30)  & 105064  & 34388.3148(30)  & 65486  & \cite{Lorenzenphysscrip1983}\\
16  & 34472.0505(30)  & 10784   & 34472.2798(30)  & 20085  & \cite{Lorenzenphysscrip1983}\\
17  & 34540.1250(30)  & 76776   & 34540.3088(30)  & -17974 &\cite{Lorenzenphysscrip1983} \\
18  & 34596.0448(30)  & 111974  & 34596.1996(30)  & 117442 & \cite{Lorenzenphysscrip1983}\\
19  &                       &         & 34642.6698(30)  & 123856 & \cite{Lorenzenphysscrip1983}\\
20  &                       &         & 34681.7220(30)  & -106   & \cite{Lorenzenphysscrip1983}\\
21  &                       &         & 34714.8646(30)  & 81996  & \cite{Lorenzenphysscrip1983}\\
22  & 34743.1426469(3)      & -1      & 34743.2226878(5)      & 9      & II \\
23  &                       &         & 34767.6811325(9)      & -18    & II\\
25  & 34807.4321004(7)      & 8       & 34807.4850300(7)      & -20    & II\\
28  & 34850.9930019(4)      & -1      & 34851.0298046(3)      & -9    &  II\\
32  & 34890.1758731(6)      & 11      & 34890.1999393(5)      & 35     & II\\
37  & 34921.6794350(10)     & -40     & 34921.6946544(4)      & -1     & II\\
50  & 34962.7480893(18)     & 54      & 34962.7540292(18)     & 55     & I\\
52  & 34966.4175066(18)     & -20     & 34966.4227658(18)     & -16    & I\\
54  & 34969.6739514(18)     & -55     & 34969.6786309(18)     & -21    & I\\
56  & 34972.5771511(18)     & 63      & 34972.5813316(18)     & 71     & I\\
58  & 34975.1764005(18)     & 24      & 34975.1801512(18)     & 39    &  I\\
60  & 34977.5126981(18)     & -17     & 34977.5160755(18)     & -18   &  I\\
60  &                       &         & 34977.5160756(18)     & -14   &  I\\
60  &                       &         & 34977.5160758(18)     & -9    &  I\\
60  &                       &         & 34977.5160758(18)     & -6    &  I\\
60  &                       &         & 34977.5160765(18)     & 13    &  I\\
60  &                       &         & 34977.5160770(5)      & 30    &  II\\
60  &                       &         & 34977.5160792(18)     & 94    &  I\\
62  & 34979.6203553(18)     & 21      & 34979.6234072(18)     & 9      & I\\
64  & 34981.5282569(18)     & -19     & 34981.5310248(18)     & -10    & I\\
66  & 34983.2608729(18)     & -55     & 34983.2633916(18)     & -15    & I\\
68  & 34984.8390359(18)     & -90     & 34984.8413326(18)     & -77    & I\\
70  & 34986.2805762(18)     & -28     & 34986.2826765(18)     & -30    & I\\
70  &                       &         & 34986.2826766(8)      & -28   &  II\\
72  & 34987.6008173(18)     & 9       & 34987.6027450(18)     & 56    &  I\\
76  & 34989.9285927(18)     & 56      & 34989.9302238(18)     & 43    &  I\\
80  & 34991.9087544(18)     & 40      & 34991.9101456(18)     & -44   &  I\\
84  & 34993.6072228(18)     & 14      & 34993.6084239(18)     & 34    &  I\\
88  & 34995.0750037(18)     & 60      & 34995.0760398(18)     & -89   &  I\\
92  & 34996.3520484(18)     & -124    & 34996.3529579(18)     & -103  &  I\\
96  & 34997.4700401(18)     & -108    & 34997.4708384(18)     & -97   &  I\\
100 & 34998.4543352(18)     & 39      & 34998.4550390(18)     & 20   & I\\
		\hline\hline
	\end{tabular}
	\endgroup
\end{table*}

In this work, 56 transitions $np_j\leftarrow4s_{1/2}$ with $n$ in the range from 22 to 100 have been measured. For the determination of the transition frequencies, the number of detected ions was recorded as a function of the laser-excitation frequency. The spectral line shape was found to be well described by a Gaussian line profile, as shown in Fig.~\ref{fig:56ptrans}(a) for the transition $22p_{3/2}\leftarrow4s_{1/2}(F=1)$.  The hyperfine splitting of $n\geq22$ Rydberg states was estimated to be below \SI{80}{kHz} from the hyperfine splitting of the $4p_{1/2}$ state and the $n^{-3}$ scaling law. Even for the lowest states measured in this study, we could neither observe a splitting nor an asymmetry in the lines. Hence we neglect the hyperfine structure in modelling the spectral line shape.

 Each transition was typically measured five times, each measurement consisting of a scan from lower to higher frequencies (up) and one from higher to lower frequencies (down) (see Fig.~\ref{fig:56ptrans}(b)). We observe a systematic shift of the transition frequency to lower (higher) transition frequencies for scans up (down) in frequency with an absolute magnitude of about \SI{20}{\kilo\hertz}. This shift is caused by delays in the determination of the laser frequency by the frequency comb, and is fully cancelled by averaging an equal number of measurements performed up and down in the laser frequency. The statistical uncertainty of the transition frequency is estimated by fitting separately Gaussian line profiles to all acquired scans and calculating the standard error of the mean of the resulting line centers. The mean transition frequency is determined from a nonlinear regression of the combined weighted data points from all individual scans, taking into account the Poissonian nature of the detection, as described in Ref.~\cite{beyerpra2018}. All measured transition wave numbers are listed in Table~\ref{ptrans}, with the uncertainties given in brackets (one standard deviation). For measurement configuration I, the \SI{50}{\kilo\hertz} uncertainty of the transition wave numbers is given by the systematic frequency-calibration error discussed in Section~\ref{Experiment}. In measurement configuration II, the statistical uncertainties are given.

\begin{figure}
	\includegraphics[trim={0 0cm 0 0},clip,width=\linewidth]{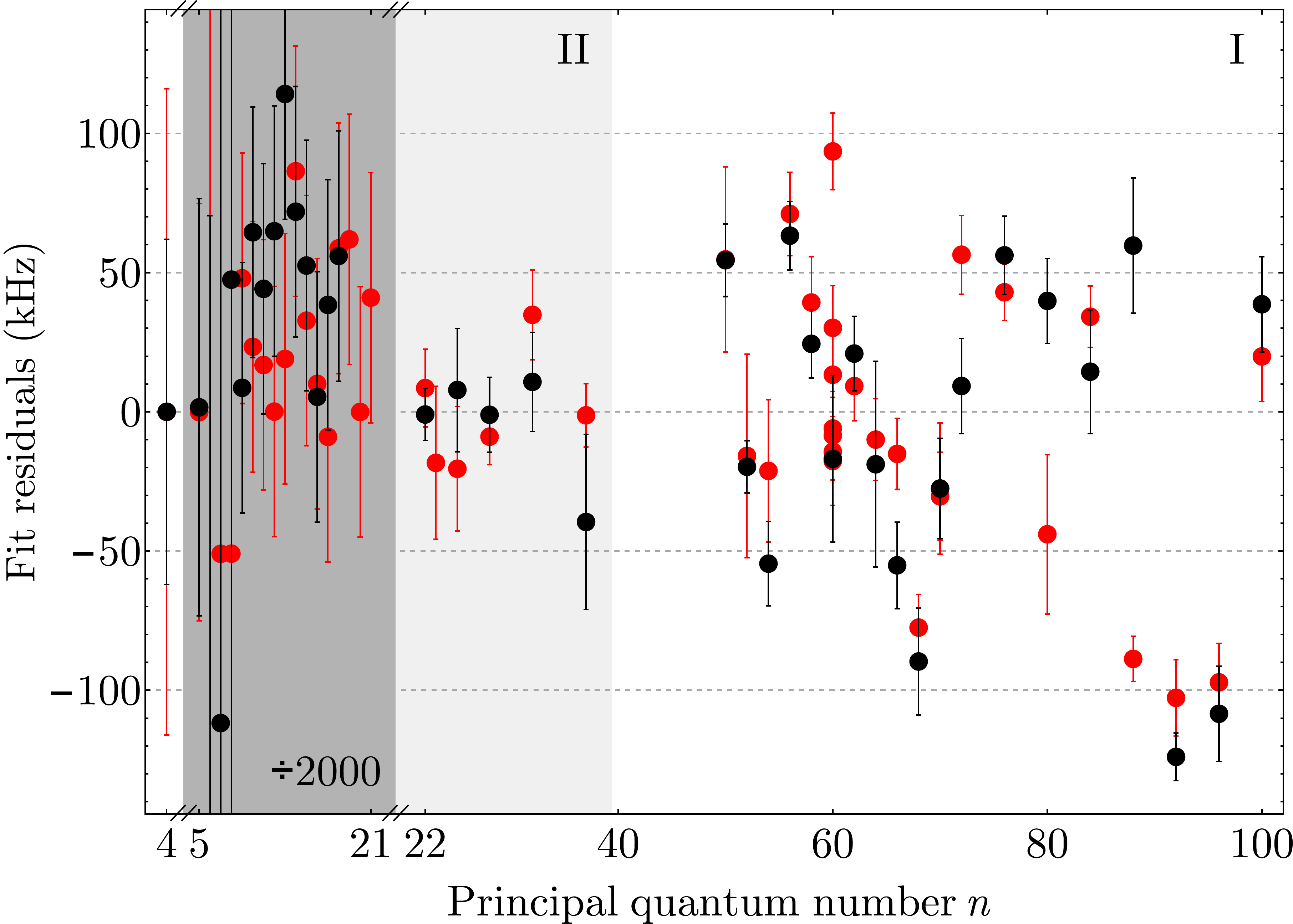}
	\caption{\label{fig:fitresiduals} Fit residuals $\delta_\mathrm{fit}$ of the global fit to the $np_{1/2}\leftarrow4s_{1/2}$ (black) and $np_{3/2}\leftarrow4s_{1/2}$ (red) series using Eqs.~(\ref{eq:rydberg}) and (\ref{eq:seriesexpansionmod}). Error bars indicate the statistical uncertainty of the transition frequencies. The fit residuals $\delta_\mathrm{fit}$ and error bars of the $np_j\leftarrow4s_{1/2}$ transition frequencies $n=5-21$ in the region with dark-gray background are divided by 2000 for clarity. The labels I and II refer to the measurement configurations discussed in Section~\ref{Experiment}.}
\end{figure}

\begin{table}
	\caption{Quantum-defect expansion coefficients and ionization wave number $E_\mathrm{I}/(hc)$ determined for the $np_{1/2}$ and $np_{3/2}$ series of $^{39}$K in a global fit of all transitions in Table~\ref{ptrans} using Eqs.~(\ref{eq:rydberg}) and (\ref{eq:seriesexpansionmod}).\label{qdp}}
	
	\begingroup\renewcommand{\arraystretch}{1}
	\begin{tabular}{c S[table-format=4.15] S[table-format=4.15]}
		\hline\hline
		\multicolumn{1}{c}{$\vphantom{\frac{\frac{a}{b}}{\frac{a}{b}}} E_\mathrm{I}/(hc)$ } & \multicolumn{2}{c}{\num{35009.8139710(22)}$_\mathrm{sys}$(3)$_\mathrm{stat}$\,\si{\per\centi\meter}}\\
		\hline
		& $\vphantom{\frac{\frac{a}{b}}{\frac{a}{b}}} np_{1/2}$ & $np_{3/2}$\\
		\hline
		$\delta_0$ & 1.71392626(9) & 1.71087854(8) \\
		$\delta_2$ & 0.23114(4) & 0.23233(4)  \\
		$\delta_4$ & 0.1948(6) & 0.1961(6) \\
		$\delta_6$ & 0.3683(23) & 0.3716(22) \\
		\hline\hline
	\end{tabular}
	\endgroup
\end{table}

For the global fit of Eq.~(\ref{eq:seriesexpansionmod}) to the term values of $np_j$ Rydberg states with $j=1/2$ and $j=3/2$, all transitions measured in this work with $n$ values between 22 and 100 are included. In addition, transitions measured in previous studies are included, covering the range of $n$ values between 4 and 21 \cite{falkepra2006,Johansson1972,Lorenzenphysscrip1983,risbergarkfys1956}. This procedure yielded a significant reduction of the correlations between the fitted parameters for the ionization energy and the expansion coefficients of the quantum defects. The two fine-structure series are fitted with a common ionization energy $E_\mathrm{I}$. A series expansion up to order 6 in Eq.~(\ref{eq:seriesexpansionmod}) was found to be sufficient to describe the energy dependence of the quantum defects. The fit parameters obtained from a weighted nonlinear regression of all transitions from Table~\ref{ptrans} are given in Table~\ref{qdp}.

The fit residuals $\delta_\mathrm{fit}$ for the individual transitions are given in Table~\ref{ptrans} and plotted in Fig.~\ref{fig:fitresiduals}. Whereas the residuals in measurement configuration II are distributed as expected for a normal distribution, the residuals in measurement configuration I are larger than expected from the statistical uncertainty of the individual transition frequencies. This additional scattering of the fit residuals was attributed to a frequency-calibration error in measurement configuration I, as explained in Section~\ref{Experiment}. The same measurement configuration was used in the determination of the Cs ionization potential \cite{deiglmayrpra2016}. In the case of Cs, the full set of individual transitions was recorded over a longer period of time, leading to a conversion of the (time-varying) systematic uncertainty into a statistical uncertainty.

The total uncertainty of the ionization energy is dominated by the frequency-calibration uncertainty of measurement set I, as discussed in Section~\ref{systematicshifts}. The obtained value for the first ionization energy of $^{39}$K with respect to the center of gravity of the ground-state hyperfine structure is \num{35009.8139710(22)}$_\mathrm{sys}$(3)$_\mathrm{stat}$\,\si{\per\centi\meter}. This value is in agreement with the most precise  previous value of \SI{35009.8140(7)}{\per\centi\meter}, reported by \citet{LORENZENoptcomm1981}, but has a 300-fold improved accuracy. The values of the quantum-defect coefficients cannot be directly compared with previous results, because different orders of the series expansion were used \cite{Lorenzenphysscrip1983}. The quoted uncertainties of the quantum-defect parameters given in Table~\ref{qdp} are the statistical uncertainties resulting from the nonlinear regression.

\begin{figure}
	\includegraphics[trim={0 0cm 0 0},clip,width=\linewidth]{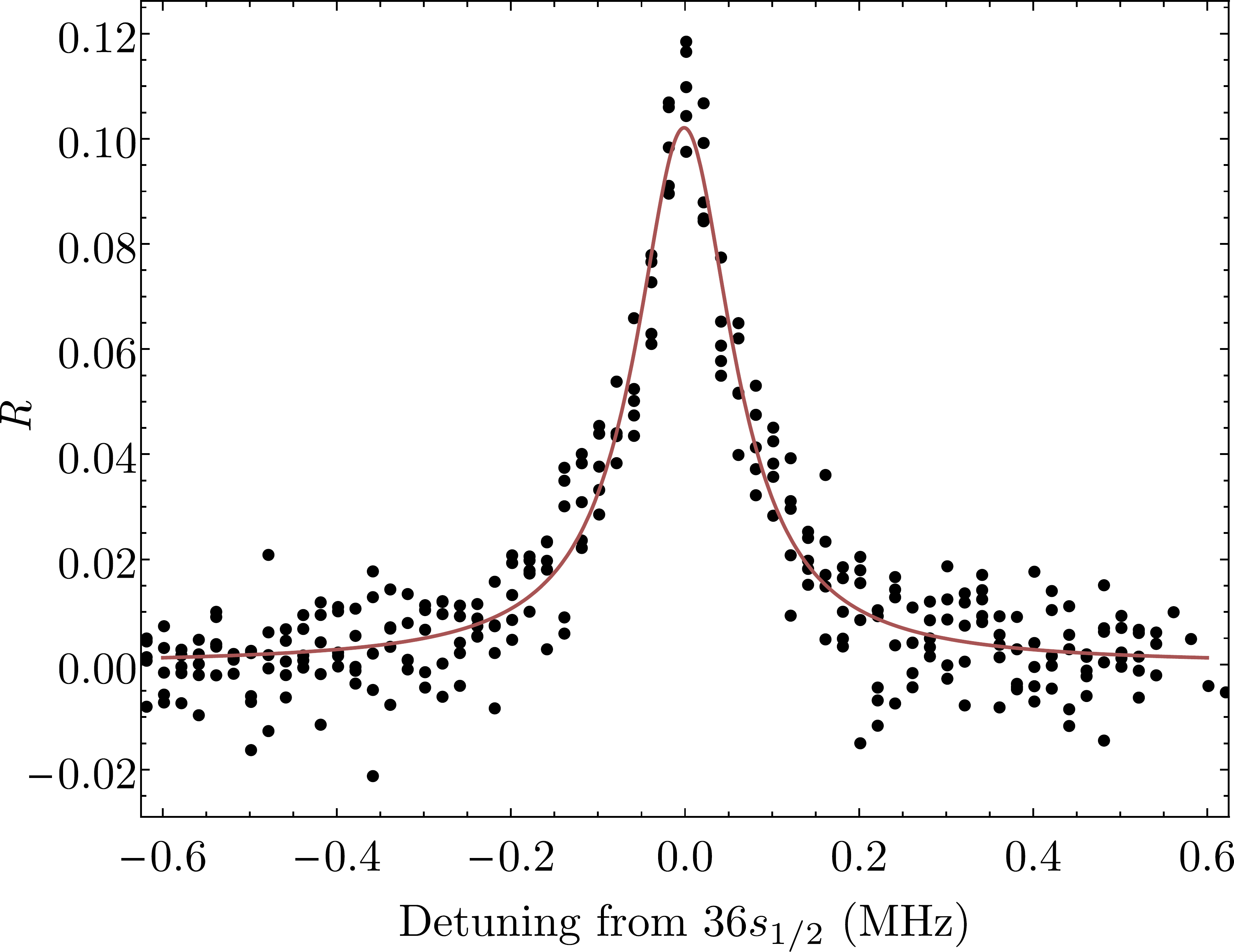}
	\caption{\label{fig:36strans} Ratio $R$ of transferred $^{39}$K atoms over total number of atoms as a function of the detuning from the $36s_{1/2}\leftarrow 35p_{3/2}$ transition frequency in MHz. The red line represents the result of the fit of a Lorentzian line profile to the observed data points (black points).}
\end{figure}

\begin{table}
	\caption{Frequencies of all measured $n's_{1/2}\leftarrow np_{3/2}$ transitions in $^{39}$K. Uncertainties reflect 1/10$^\textrm{th}$ of the linewidth. The last column gives the residuals $\delta_\mathrm{fit}$ obtained in the fit of the quantum-defect parameters. \label{strans}}
	\begingroup\renewcommand{\arraystretch}{1.1}
	\begin{tabular}{c S[table-format=8.10] S[table-format=2.0]}
		\hline\hline
		\multicolumn{1}{c}{Transition $f \leftarrow i$} & \multicolumn{1}{c}{$\nu$ (MHz)} &  \multicolumn{1}{c}{$\delta_\mathrm{fit}$ (kHz)} \\
		\hline
		$35s_{1/2}\leftarrow 35p_{3/2}$ & 85498.564(19) & -11  \\
		$36s_{1/2}\leftarrow 35p_{3/2}$ & 92450.458(14) & -13 \\
		$38s_{1/2}\leftarrow 37p_{3/2}$ & 77707.357(30) & -11 \\
		$50s_{1/2}\leftarrow 51p_{3/2}$ & 84493.027(20)  & 14 \\
		$53s_{1/2}\leftarrow 51p_{3/2}$ & 80347.017(10)  & 11 \\
		$67s_{1/2}\leftarrow 70p_{3/2}$ & 77535.721(20) & 10 \\
		$75s_{1/2}\leftarrow 70p_{3/2}$ & 85052.474(19) & -1  \\
		\hline\hline
	\end{tabular}
	\endgroup
\end{table}

\begin{table*}
	\caption{Frequencies of all measured $n'd_j\leftarrow np_{3/2}$ transitions in $^{39}$K.  Uncertainties reflect 1/10$^\textrm{th}$ of the linewidth. The third and last columns give the residuals $\delta_\mathrm{fit}$ obtained in the fit of the quantum-defect parameters.\label{dtrans}}
	\begingroup\renewcommand{\arraystretch}{1}
	\begin{tabular}{c S[table-format=8.10] S[table-format=2.0] S[table-format=8.10] S[table-format=2.0]}
		\hline\hline
		\multicolumn{1}{c}{Transition $\vphantom{\frac{\frac{a}{b}}{\frac{a}{b}}} f\leftarrow i$}& \multicolumn{1}{c}{$\nu_{d_{3/2}}$ (MHz)} &  \multicolumn{1}{c}{$\delta_\mathrm{fit}$ (kHz)} & \multicolumn{1}{c}{ $\nu_{d_{5/2}}$ (MHz)} & \multicolumn{1}{c}{ $\delta_\mathrm{fit}$ (kHz)}\\
		\hline
		$32d_j\leftarrow34p_{3/2}$ & 113369.342(11) &  -8 & 113405.206(11) & 8 \\
		$33d_j\leftarrow35p_{3/2}$ & 103387.013(12) &  15 & 103419.729(12) &  -33\\
		$35d_j\leftarrow37p_{3/2}$ & 86678.837(16) &  14 & 86706.135(14)& 24 \\
		$48d_j\leftarrow51p_{3/2}$ & 90302.892(14) & 3 & 90313.359(10) &  8\\
		$51d_j\leftarrow51p_{3/2}$&  75504.404(14) &  -26 & 75495.704(9) &  -9\\
		\hline\hline
	\end{tabular}
	\endgroup
\end{table*}

\section{Millimeter-wave spectroscopy and $s$-, $d$-, $f$-, and $g$-series quantum defects}
\label{mmwavespec}
The quantum defects of other Rydberg series of $^{39}$K were determined by millimeter-wave spectroscopy. Transitions to $ns_{1/2}$, $nd_j$, $nf_j$ and $ng_j$ are driven by radiation in the range from 18 to \SI{250}{\giga\hertz}: $i)$ $ns_{1/2}$ and $nd_j$ states are populated by a one-photon transition from $n' p$ states, $ii)$ $nf_j$ states are populated either by a two-photon transition from $n' p_{j}$ states or by a one-photon transition from $n' d_{j}$ states and $iii)$ $ng_j$ states are populated by a one-photon transition from $n' f_{j}$ states, which have been populated by a two-photon transition from $n''p_{3/2}$ states. The optical and all millimeter-wave fields are applied sequentially to the samples of $np$ Rydberg atoms and do not overlap in time.

The spectral line shapes of the transitions to \textit{s, d, f} and \textit{g} Rydberg states (Tables~\ref{strans}-\ref{gtrans}) are well described by Lorentzian line profiles, as depicted in Fig.~{\ref{fig:36strans}} for the transition $36s_{1/2}\leftarrow 35p_{3/2}$, and are only slightly broader than expected for the Fourier-transform limit of rectangular pulses. The observed relative transition strengths to the two $d_j$ fine-structure components confirm an inverted fine structure for $nd_j$ Rydberg states ($n\geq 32$) \cite{LORENZENoptcomm1981}. All transitions $n's_{1/2} \leftarrow np_{3/2}$ ($35 \leq n' \leq 75$) and $n'd_{j} \leftarrow np_{3/2}$ ($32 \leq n' \leq 51$) observed in this work are listed in Tables~\ref{strans} and \ref{dtrans}. For the determination of the quantum-defect parameters for these series, the absolute energies of the final states were obtained by combining the measured transition energies with the energy of the initial $np$ state calculated from Eqs.~(\ref{eq:rydberg}) and (\ref{eq:seriesexpansionmod}) and the parameters listed in Table~\ref{qdp}. The resulting sets of term values for the $s_{1/2}$ and $d_j$ series were augmented by the term values of \citet{LORENZENoptcomm1981} and \citet{Stalnaker17} for lower values of $n$. The fit to the term values of the $s_{1/2}$ series additionally included the term value of the $4s_{1/2}$ state (\SI{0}{\per\centi\meter}). The quantum-defect parameters were then obtained from separate, weighted fits of Eq.~\eqref{eq:rydberg} (with $E_\mathrm{I}$ fixed to the value determined from the $p$ series) and Eq.~\eqref{eq:seriesexpansionmod} to the three sets of term values. The weights for the term values used in the fits were obtained from the combined uncertainty from initial and final states, whereas the quoted uncertainties for the term values from Ref.~\cite{LORENZENoptcomm1981} and \cite{Stalnaker17} were used. The uncertainty of the 4$s_{1/2}$ term value was chosen to be \SI{67}{kHz}, in accordance with the uncertainty of the ionization energy obtained in the extrapolation of the $np_j$ Rydberg series. The resulting quantum-defect parameters are given in Table~\ref{qdsd}.

\begin{figure*}
    \includegraphics[trim={0 0cm 0 0},clip,width=\linewidth]{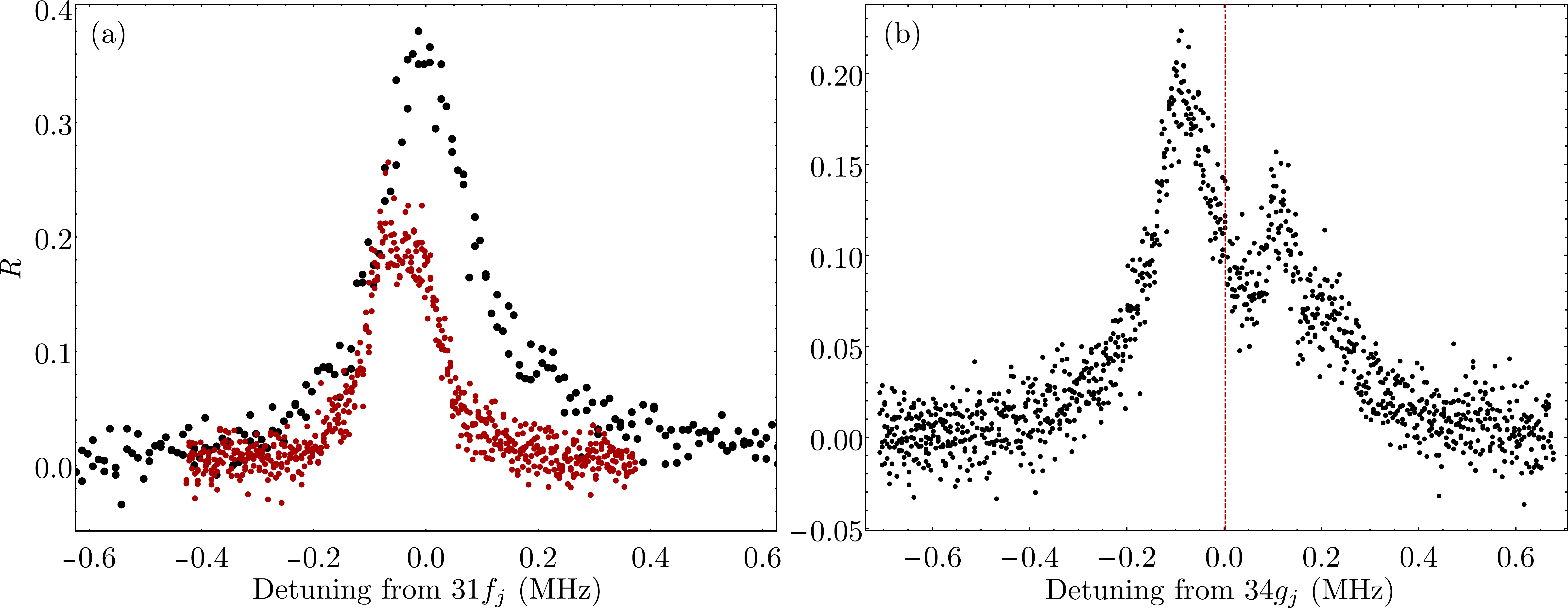}
	\caption{\label{fig:34gtrans} (a)  Ratio $R$ of transferred $^{39}$K atoms over total number of atoms as a function of the detuning from the $31f_{j}$ state for the transition $31f_j\leftarrow32d_{5/2}$ (black) and the transition $31f_j\leftarrow32d_{3/2}$ (red).  (b) Ratio $R$ of transferred $^{39}$K atoms over total number of atoms as a function of the detuning from the $34g_{j}\leftarrow 35f_{j}$ transition frequency in MHz. The center of gravity of the black data points is given as a red line.}
\end{figure*}

An example of a one-photon transition from $n d_j$ to $n' f_j$ is given in Fig. \ref{fig:34gtrans}\,(a). Even for the lowest investigated state, $30f$, the two fine-structure components are not resolved in our measurements. Because the fine-structure splitting has contributions from the spin-orbit interaction in a pure Coulomb interaction potential (the hydrogenic fine-structure splitting, \SI{490}{kHz} for $31f$ \cite{gallagher1994}) and the interaction between Rydberg and core electrons, this near-degeneracy of the fine-structure components indicates that both contributions are of comparable magnitude and opposite sign for $l=3$ in potassium. In order to determine an upper bound for the fine-structure splitting, we drive the one-photon transitions $31f_j\leftarrow32d_{3/2}$ and $31f_j\leftarrow32d_{5/2}$ from the two fine-structure components of the $32d$ state. Selection rules restrict transitions from $32d_{3/2}$ to the $j=5/2$ component of $31f$, whereas both the $31f_{5/2}$ and $31f_{7/2}$ components are accessible from 32$d_{5/2}$. Spectra of the two transitions, corrected for the experimentally determined fine-structure splitting of the $32d$ state, are compared in Fig. \ref{fig:34gtrans}~(a). The line originating from $32d_{5/2}$ exhibits a broadening towards higher transition frequencies, indicating an inverted fine structure of the $31f$ state, which lies energetically below the $32d$ state. Taking into account the uncertainty of the $32d$ fine-structure interval, we can exclude a splitting larger than \SI{100}{kHz}, which is in qualitative agreement with calculations by Pyper and Marketos \cite{pyper81}, who predict the fine-structure of the $30f$ state to be inverted and to be less than half of the hydrogenic fine-structure splitting.

The quantum-defect parameters for the center of gravity of the $n f_j$ states are obtained from the experimental transition frequencies (see Table~\ref{ftrans}) in a fit based on  Eqs.~\eqref{eq:rydberg} and \eqref{eq:seriesexpansionmod}. The quantum-defect parameters of the initial $n p_j$ and $n d_j$ states are fixed to the values in Table~\ref{qdp} and Table~\ref{qdsd}, respectively. Because of the smaller range of $n$ values fitted in this series, two expansion coefficients suffice to describe the energy dependence of the quantum defects. The fit residuals, also given in Table~\ref{ftrans}, show much larger deviations than those obtained for the  $s_{1/2}$ and $d_j$ series (Tables~\ref{strans} and \ref{dtrans}), which we attribute to uncertainties in the determination of the line centers when fitting a single Lorentzian line profile to a resonance with unresolved fine structure.

Transitions to the $g$ series show a fine-structure splitting which is comparable to the hydrogenic fine-structure splitting of \SI{223}{kHz} in the case of the $34g$ state (see Fig.~\ref{fig:34gtrans}\,(b)). The observed line shape is, however, very sensitive to small variations in the applied electric fields and to the power of the millimeter-wave radiation, even at the strongest attenuation for which transitions can still be recorded with sufficient signal-to-noise ratio. The line positions are thus determined from the center of gravity of the observed overall line shape, indicated in red in Fig.~\ref{fig:34gtrans}\,(b) for the $34g$ state. The quantum-defect parameters are then extracted similarly to the other series, where the quantum defects of the initial $n f_j$ states are fixed to the values of Table~\ref{qdsd}.

\begin{table}
	\caption{Frequencies of all measured $n'f_j\leftarrow np_{j}$ and $n'f_j\leftarrow nd_j$ transitions in $^{39}$K. Uncertainties reflect 1/10$^\textrm{th}$ of the linewidth. The last column gives the residuals $\delta_\mathrm{fit}$ obtained in the fit of the quantum-defect parameters. \label{ftrans}}
	\begingroup\renewcommand{\arraystretch}{1.1}
	\begin{tabular}{c S[table-format=8.10] S[table-format=2.0]}
		\hline\hline
		\multicolumn{1}{c}{Transition $f\leftarrow i$} & \multicolumn{1}{c}{$\nu$ (MHz)} &  \multicolumn{1}{c}{$\delta_\mathrm{fit}$ (kHz)} \\
		\hline
		$30f_j\leftarrow32p_{3/2}$ & 71683.751(34) & -36 \\
		$31f_j\leftarrow32d_{3/2}$ & 156556.214(20)  & 32  \\
		$31f_j\leftarrow32d_{5/2}$ & 156520.393(14)  & -26 \\
		$32f_j\leftarrow34p_{3/2}$ & 59123.780(5)    & -38 \\
		$35f_j\leftarrow37p_{1/2}$ & 44786.853(6)   & 92  \\
		$35f_j\leftarrow37p_{3/2}$ & 45243.104(8)   & 71  \\
		$38f_j\leftarrow40p_{3/2}$ & 35388.501(6)    & -31 \\
		$38f_j\leftarrow40p_{3/2}$ & 35388.553(8)   & -83 \\
		$39f_j\leftarrow41p_{1/2}$ & 32415.165(5)   & -67 \\
		$39f_j\leftarrow41p_{3/2}$ & 32745.659(12)   & 42  \\
		$40f_j\leftarrow42p_{3/2}$  & 30359.580(9)   & 58  \\
		$41f_j\leftarrow43p_{3/2}$ & 28199.945(7)    & -86 \\
		$42f_j\leftarrow44p_{3/2}$ & 26240.076(15)   & 86  \\
		$45f_j\leftarrow47p_{3/2}$ & 21349.724(13)   & 5   \\
		$47f_j\leftarrow49p_{3/2}$  & 18746.660(11)    & -19\\
		\hline\hline
	\end{tabular}
	\endgroup
\end{table}

\begin{table}
	\caption{Frequencies of all measured $n'g_{j}\leftarrow nf_{j}$ transitions in $^{39}$K. Uncertainties reflect 1/10$^\textrm{th}$ of the linewidth. The last column gives the residuals $\delta_\mathrm{fit}$ obtained in the fit of the quantum-defect parameters.\label{gtrans}}
	\begingroup\renewcommand{\arraystretch}{1.1}
	\begin{tabular}{c S[table-format=8.10] S[table-format=2.0]}
		\hline\hline
		\multicolumn{1}{c}{Transition $f\leftarrow i$}& \multicolumn{1}{c}{$\nu$ (MHz)} &  \multicolumn{1}{c}{$\delta_\mathrm{fit}$ (kHz)} \\
		\hline
		$30g_j\leftarrow31f_{j}$ & 230525.74(6) &  -43  \\			
		$34g_j\leftarrow35f_{j}$ & 159250.24(5) &  39  \\
		$35g_j\leftarrow36f_{j}$ & 146162.36(4) &  73  \\
		$37g_j\leftarrow38f_{j}$ & 123993.16(5) &  -169  \\
		\hline\hline
	\end{tabular}
	\endgroup
\end{table}

\begin{table*}
\caption{Quantum-defect expansion coefficients for the $nl_j$ series obtained from a fit of \equationref{eq:rydberg} and \equationref{eq:seriesexpansionmod} to the transition frequencies given in Table~\ref{strans}-\ref{gtrans} using the value of $E_\mathrm{I}$ obtained from extrapolation of the $np_{j}$ series.\label{qdsd}}
\begin{tabular}{c S[table-format=4.15] S[table-format=3.10] S[table-format=3.7] S[table-format=3.7]}
	\hline\hline
 & \multicolumn{1}{c}{$\delta_0$} & \multicolumn{1}{c}{$\delta_2$ }& \multicolumn{1}{c}{$\delta_4$ }& \multicolumn{1}{c}{$\delta_6$ }\\
\hline
$s_{1/2}$ & 2.18020826(5) & 0.134534(17) & 0.0952(3) & 0.0021(8)  \\
$d_{3/2}$ & 0.27698453(19) & -1.02691(24) & -0.665(27)& 10.9(8) \\
$d_{5/2}$ & 0.27715665(14) &  -1.02493(17) & -0.640(20)& 10.0(6)\\
$f_{j}$ & 0.0094576(6) &  -0.0446(6)& &  \\
$g_{j}$ & 0.0024080(25) &  -0.0209(27)& & \\
\hline\hline
\end{tabular}
\end{table*}

\section{Ground state hyperfine splitting}
\label{gssplitting}
Measurement of the  $F=2\leftarrow F=1$ transition by RF spectroscopy yields a central line position of \SI{461.719700(20)}{MHz}, as explained in Section~\ref{Experiment}. The main uncertainty of the transition frequency arises from the width of the line.

\begin{figure}
	\includegraphics[trim={0 0cm 0 0cm},clip,width=\linewidth]{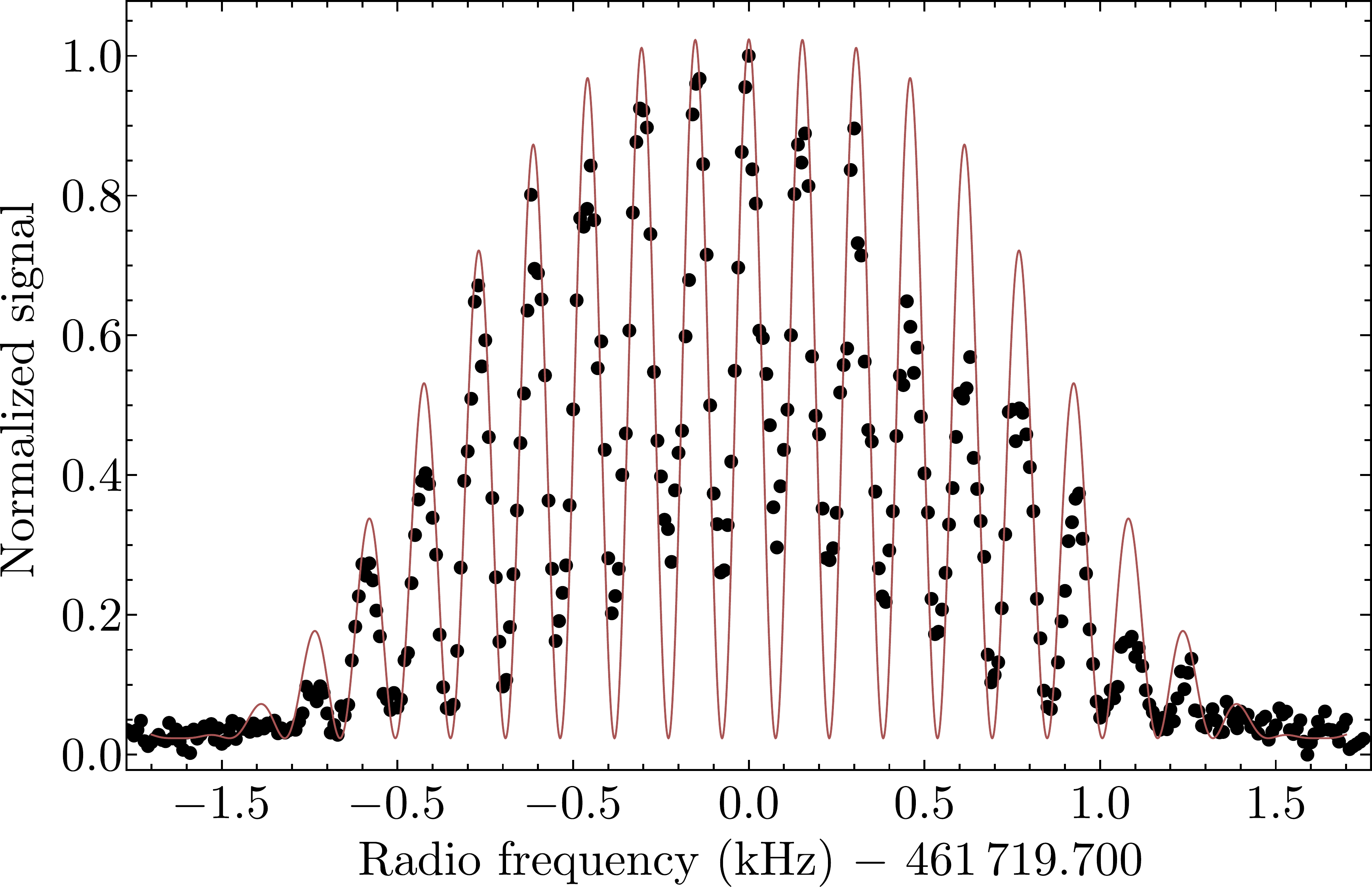}
	\caption{\label{fig:hyperfinetrans}(Black points) Population in the $F=2$ component of the 4$s_{1/2}$ ground state of $^{39}$K as a function of the applied RF frequency in a Ramsey sequence with \SI{5.8}{ms} interpulse delay (see text for details). (Red curve) Fit of Eq. (1) of Ref.~\cite{truppe2013search} to the experimental data.}
		\label{ramseyfit}
\end{figure}

For a reduction of the measurement error, a Ramsey pulse sequence is employed. This sequence ($\frac{\pi}{2}-\tau_\mathrm{d}-\frac{\pi}{2}$) consists of two $\frac{\pi}{2}$ pulses separated by a field-free evolution period of length $\tau_\mathrm{d}$. Monitoring the transferred population as a function of the RF frequency yields an interference pattern known as Ramsey fringes. Under typical experimental conditions, the pulse length for a $\frac{\pi}{2}$ pulse is determined to be \SI{600}{\micro s} by driving Rabi cycles between the two hyperfine components. The Ramsey-interferometry spectrum for the measurement with an interpulse delay of \SI{5.8}{ms} is fitted with the expression given in Ref.~\cite{truppe2013search}, yielding a central frequency of \SI{461.719700(5)}{MHz} (see Fig.~\ref{ramseyfit}). We estimate the uncertainty of the transition frequency by comparing the center frequency of Ramsey fringes for varying interpulse delays $\tau_\mathrm{d}$. Note that the experimental upper bound on the residual magnetic field of \SI{7}{mG} (see Section \ref{Experiment}) limits the possible residual Zeeman shift of the $F=2,m_F=0 \leftarrow F=1,m_F=0$ transition to \SI{1}{Hz}. Our experimental results are compared with other experiments in Table~\ref{tab:results:hyperfine}. The values obtained for the hyperfine splitting of the $4s_{1/2}$ state in $^{39}$K are in agreement with the values reported in Refs.~\cite{bloom,dahmen67}, but differ significantly from the more precise values reported in Refs.~\cite{beckmann74,chan70,arias2019}.  The discrepancy between our value and the value reported by \citet{arias2019} is almost \SI{50}{Hz} or 10\,$\sigma$ and can be mostly explained by the quadratic Zeeman shift resulting from the presence of a magnetic background field of about \SI{60}{mG} in the latter experiment, as explained in Ref.~\cite{arias2019}. Reference \cite{AntoniMicollier2017} reports an all-optical measurement of the hyperfine splitting with low statistical uncertainty, but uncontrolled systematical errors.

\begin{table}
	\centering
	\caption[]{Comparison of ground-state hyperfine splitting of $^{39}$K obtained in this work by radio-frequency (RF) and Ramsey-interferometry spectroscopy (Ramsey interferometry) with literature values. The literature values are reviewed in \cite{arimondo77}. The abbreviation ABMR stands for atomic beam magnetic resonance.}
	\label{tab:results:hyperfine}
	\begin{tabular}{l S[table-format=4.8]}
		\hline\hline
		Method&\multicolumn{1}{c}{Hyperfine splitting (\si{MHz})}\\
		\hline
        RF (this work)	&461.719700(20)\\
		Ramsey interferometry (this work)	& 461.719700(5)\\
		Optical pumping \cite{bloom}			&461.719690(30)\\
		ABMR	\cite{dahmen67}					&461.719723(38)\\
		ABMR			\cite{beckmann74}		&461.7197202(14)\\
		ABMR		\cite{chan70}				&461.7197201(6)\\
        Ramsey interferometry \cite{arias2019} & 461.7197480(6)  \\
		\hline\hline
	\end{tabular}
\end{table}

\section{The polarizability of the K$^+$ core}
\label{corepolarization}
Polarizabilities of atoms and ions have been studied extensively both experimentally and theoretically over the last decades (see, e.g., Ref.~\cite{Mitroy2010} and references therein). Their exact values are important in studies of atom-atom pair interactions \cite{marinescu1994}, neutral-ion collisions \cite{allmendinger2016observation}, and in the evaluation of the uncertainties of atomic \cite{bloom2014} and ionic clock transitions \cite{Rosenband1808}. The polarizability of the ion core even constitutes the largest uncertainty in a proposed measurement of the Rydberg constant using circular Rydberg states~\cite{Raithelrydbergconstant}. Polarizabilities have been determined previously using refractometry of salt solutions \cite{Fajans1924} and crystals \cite{wilson1970,tessmann1953}. The polarizabilities of free ions can be determined from spectra of atoms as proposed by Born and Heisenberg \cite{Born1924}, Freeman and Kleppner \cite{freemankleppnerpra1976},  Mayer and Mayer \cite{MayerMayerphysrev1933} and \citet{sansonetti1981polarization}. In the analysis of our data we follow the procedure used by \citet{sansonetti1981polarization}.

We assume that the energy of a given Rydberg state $E(n,l)$ can be calculated \textit{ab-initio} as a sum of individual contributions,
\begin{equation}
\begin{split}
E(n,l)&=E_\mathrm{I}-\frac{hcR_\mathrm{K}}{{n}^2}-\Delta E_\mathrm{rel}(n,l)-\Delta E_\mathrm{pol}(n,l)\\&-\Delta E_\mathrm{pen}(n,l)
-\Delta E_\mathrm{xch}(n,l)
\end{split}\label{eq:QDcontributions}
\end{equation}
where $E_\mathrm{I}$ is the ionization energy, $n$ is the principal quantum number, $\Delta E_\mathrm{rel}(n,l)$ is the relativistic energy correction, $\Delta E_\mathrm{pol}(n,l)$ is the energy shift caused by the polarization of the ion core by the valence electron, $\Delta E_\mathrm{pen}(n,l)$ is the energy shift resulting from the valence electron penetrating into the ion core and $\Delta E_\mathrm{xch}(n,l)$ is the energy shift caused by exchange interaction between the valence electron and all core electrons. A physical relation between the terms of Eq.~\ref{eq:QDcontributions} and the expansion coefficients in Eq.~\eqref{eq:seriesexpansion} is discussed by \citet{drakeswainsonpra1991}. Further corrections, such as radiative energy corrections \cite{Ginges2016}, are neglected.

For Rydberg states with orbital angular-momentum quantum number $l\geq3$, the deviation from hydrogenic behavior mainly originates from the polarization of the ion core by the Rydberg electron. Assuming the validity of \equationref{eq:QDcontributions}, one can determine any one of the energy contributions if all others are known.

\subsection{Polarization formula and corrections}

\subsubsection{Penetration, exchange and relativistic effects}
\label{calc}
Penetration, exchange and hydrogenic relativistic effects were calculated as described in Ref.~\cite{sansonetti1981polarization}. The one-electron wave functions of the valence electron and the core electrons used in the analysis were obtained from the atomic-structure code by Cowan \cite{cowan1981theory}. This code performs a relativistically modified Hartree-Fock (HFR) calculation and was modified to output the one-electron wave functions of multi-electron atoms. The relativistic energy correction $\Delta E_\mathrm{rel}(n,l)$ is taken to be equal to  the relativistic kinetic-energy correction for a pure Coulomb field. Non-hydrogenic corrections are smaller than \SI{50}{kHz} for all states considered in this paper and are hence neglected. The corresponding values are given in Table~\ref{comparisiontheoryexp}.

\begin{table*}
	\caption{Calculated polarization $\Delta E_\mathrm{pol}/h$, exchange $\Delta E_\mathrm{xch}/h$, penetration $\Delta E_\mathrm{pen}/h$ and hydrogenic relativistic $\Delta E_\mathrm{rel}/h$ contributions to the term-values of the selected states of the $f$ and $g$ series. See text for details. \label{comparisiontheoryexp}}

		\begin{tabular}{c S[table-format=8.7] S[table-format=8.7] S[table-format=8.7] S[table-format=8.7]}
			\hline\hline
			\multicolumn{1}{c}{State} & \multicolumn{1}{c}{$\Delta E_\mathrm{pol}/h$ (MHz)} & \multicolumn{1}{c}{$\Delta E_\mathrm{xch}/h$ (MHz)} & \multicolumn{1}{c}{$\Delta E_\mathrm{pen}/h$ (MHz)} & \multicolumn{1}{c}{$\Delta E_\mathrm{rel}/h$ (MHz)}\\
			\hline
$30f$ & 2199.84 & 56.17 & 36.01 & 1.69 \\
$31f$ & 1994.32 & 50.94 & 32.66 & 1.54 \\
$32f$ & 1813.61 & 46.35 & 29.72 & 1.40 \\
$35f$ & 1387.00 & 35.49 & 22.76 & 1.08 \\
$38f$ & 1084.30 & 27.78 & 17.81 & 0.85 \\
$39f$ & 1003.16 & 25.70 & 16.48 & 0.79 \\
$40f$ & 929.91  & 23.83 & 15.28 & 0.73 \\
$41f$ & 863.62  & 22.14 & 14.20 & 0.68 \\
$42f$ & 803.48  & 20.60 & 13.21 & 0.63 \\
$45f$ & 653.45  & 16.77 & 10.75 & 0.52 \\
$47f$ & 573.63  & 14.72 & 9.44  & 0.46 \\
\\[-1ex]
$30g$ & 579.60 & 0.33 & 0.18 & 1.28 \\
$34g$ & 398.81 & 0.23 & 0.12 & 0.89 \\
$35g$ & 365.71 & 0.21 & 0.11 & 0.82 \\
$37g$ & 309.73 & 0.18 & 0.10 & 0.70 \\
\hline\hline
		\end{tabular}
\end{table*}

\subsubsection{Polarization}
When treating the polarization of the ion core in a multipole expansion up to quadrupole contributions, the energy shift caused by the polarizability of the ion core $\Delta E_\mathrm{pol}(n,l)$ is given in perturbation theory as
\begin{equation}
\label{eq:energy}
\begin{split}
	\Delta E_\mathrm{pol}(n,l)&=hcR_\mathrm{K}a_0\\
	&\left(\alpha'_\mathrm{d}\left\langle R^{-4}(n,l)\right\rangle+\alpha'_\mathrm{q}\left\langle R^{-6}(n,l)\right\rangle\right)\;,
	\end{split}
\end{equation}
where $\alpha'_\mathrm{d}$ and $\alpha'_\mathrm{q}$ are the effective dipole and quadrupole polarizability volumes, respectively, and $\left\langle R^{i}(n,l) \right\rangle$ are the expectation values of $R^{i}$, calculated with the hydrogen wave functions \cite{Bockasten1974}. To obtain the static polarizability volumes $\alpha_\mathrm{d}$ and $\alpha_\mathrm{q}$ from the effective polarizabilities, one has to account for nonadiabatic contributions arising from the motion of the valence electron \cite{eissa1967polarization}  (neglecting terms with $i<-6$)
\begin{widetext}
\begin{equation}
\Delta E_\mathrm{pol}(n,l)=hcR_\mathrm{K}a_0\\
\left[\alpha_\mathrm{d}\left(y_0^\mathrm{d}(n,l)\left\langle R^{-4}(n,l)\right\rangle + y_2^\mathrm{d}(n,l)\left\langle R^{-6}(n,l)\right\rangle\right)+\alpha_\mathrm{q}y_0^\mathrm{q}(n,l)\left\langle R^{-6}(n,l)\right\rangle\right]\;,
\label{polfit}
\end{equation}
\end{widetext}
where $y_i^j$ are the expansion coefficients defined by Eissa and \"Opik \cite{eissa1967polarization}. In the adiabatic approximation, $y_0^j=1$ and $y_2^j=0$. To obtain the nonadiabatic expansion coefficients, the second-order perturbation energy given by Eq.~(28) in Ref.~\cite{eissa1967polarization} is minimized. This minimization requires prior knowledge of the static dipole and quadrupole polarizabilities, which have therefore to be determined in an iterative approach, as described below. The values of $y_i^j$ are in general $n$ and $l$ dependent, but were found to be constant for a given $l$ series in the $n$ range of our measurements, as given in Table~\ref{nonadiabatic}. The final parameters used for the calculation of $y_i^j$ are given in Table~\ref{paramnonadiabatic}. The values of $\alpha_\mathrm{d}$ and $\alpha_\mathrm{q}$ given in this table constitute the final result of our analysis.

\begin{table}
	\caption{Nonadiabatic expansion coefficients $y_i^j$ obtained from Eq.~(28) of Ref.~\cite{eissa1967polarization} using one-electron wave functions \cite{cowan1981theory}. \label{nonadiabatic}}
	\begin{ruledtabular}
		\begingroup\renewcommand{\arraystretch}{1}
		\begin{tabular}{ccccc}
			&\multicolumn{2}{c}{dipole}&\multicolumn{2}{c}{quadrupole}\\
			\hline
			Series & $\vphantom{\frac{\frac{a}{b}}{\frac{a}{b}}} y_0^\mathrm{d}$ & $y_2^\mathrm{d}$ $\left(a_0^2\right)$& $y_0^\mathrm{q}$&$y_2^\mathrm{q}$ $\left(a_0^2\right)$\\
			\hline
			$f$ & 1.026 & $-2.764$ & 0.978 & $-3.211$ \\
			$g$ & 1.018 & $-4.213$ & 1.000 & $-4.959$ \\
			
		\end{tabular}
		\endgroup
	\end{ruledtabular}
\end{table}

\begin{table*}
	\caption{Parameters used in the calculation of the expansion coefficients given in Table~\ref{nonadiabatic}. $R_0^l$ are the lower integration bounds, $\left\langle X_j|X_j\right\rangle$ are obtained from oscillator sum rules, and $\alpha_\mathrm{d}$ and $\alpha_\mathrm{q}$ are the dipole and quadrupole polarizabilities. \label{paramnonadiabatic}}
	\begin{ruledtabular}
		\begingroup\renewcommand{\arraystretch}{1}
		\begin{tabular}{cccccc}

			 $\vphantom{\frac{\frac{a}{b}}{\frac{a}{b}}} R_0^f$ $\left(a_0\right)$ & $R_0^g$  $\left(a_0\right)$ & $\left\langle X_1|X_1\right\rangle$ $\left(a_0^5\right)$& $\left\langle X_2|X_2\right\rangle$ $\left(a_0^7\right)$ & $\alpha_\mathrm{d}$ $\left(a_0^3\right)$ &$\alpha_\mathrm{q}$ $\left(a_0^5\right)$\\
			\hline
			 2.20 & 2.64 & 2.48 & 8.11 & 5.4880 & 17.89 \\
			
		\end{tabular}
		\endgroup
	\end{ruledtabular}
\end{table*}

\subsection{Determination of dipole and quadrupole polarizabilities}

\begin{figure}
	\includegraphics[trim={0 0cm 0 0},clip,width=\linewidth]{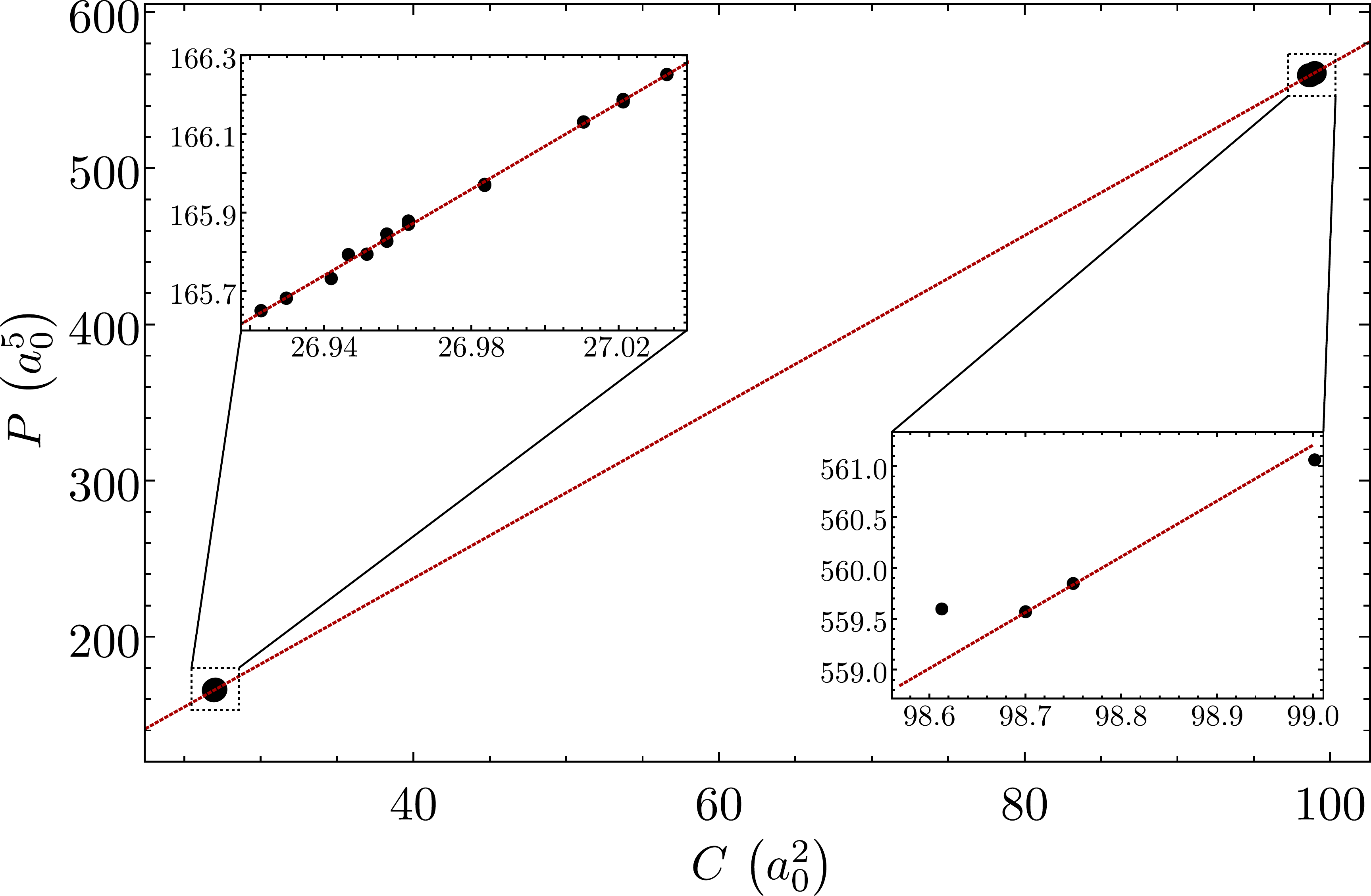}
	\caption{\label{fig:fit} Simultaneous fit of a linearization of the polarization formula (Eq.~(\ref{polfit})) (red) to the $f$ and $g$ series of $^{39}$K (black dots).}
\end{figure}

From calculated exchange, penetration and relativistic corrections, the ionization energy $E_\mathrm{I}$ and the energies of a set of states $E(n,l)$, the energy contribution of the ion-core polarizability can be determined directly from Eqs.~(\ref{eq:QDcontributions}) and (\ref{eq:energy}). A linearization of Eq.~(\ref{polfit})
\begin{equation}
	P=\alpha_\mathrm{d}C+\alpha_\mathrm{q},
\end{equation}
with
\begin{equation}
	P=\frac{\Delta E_\mathrm{pol}}{hcR_\mathrm{K}a_0y_0^\mathrm{q}\big\langle R^{-6}\big\rangle}
\end{equation}
and
\begin{equation}
	C=\frac{y_0^\mathrm{d}\big\langle R^{-4}\big\rangle+y_2^\mathrm{d}\big\langle R^{-6}\big\rangle}{y_0^\mathrm{q}\big\langle R^{-6}\big\rangle}
\end{equation}
allows us to extract the static dipole and quadrupole polarizabilities $\alpha_\mathrm{d}$ and $\alpha_\mathrm{q}$ from the  gradient and intercept, as illustrated in Fig.~\ref{fig:fit} for the $f$ and $g$ series of $^{39}$K. The linearity is verified over the range of $n$ values for which we have obtained experimental data.

As stated above, the expression used for the determination of the nonadiabatic expansion coefficients depends itself on the dipole and quadrupole polarizabilities $\alpha_\mathrm{d}$ and $\alpha_\mathrm{q}$. Repeated calculation of the expansion coefficients $y_i^j$ and fits based on the polarization formula (\ref{polfit}) reached convergence after a few iterations only, independent of the initial value of the polarizabilities.

\begin{table}
	\caption{Comparison of the dipole $\alpha_\mathrm{d}$ and quadrupole $\alpha_\mathrm{q}$ polarizability of the K$^+$-ion core obtained in this work with previous theoretical \cite{mahan1980,lim2002} and experimental \cite{oepik,eissa1967polarization,risbergarkfys1956} results.\label{poldq}}
	\begingroup\renewcommand{\arraystretch}{1}
	\begin{tabular}{l S[table-format=5.8] S[table-format=7.8]}
		\hline\hline
		& \multicolumn{1}{c}{$\vphantom{\frac{\frac{a}{b}}{\frac{a}{b}}} \alpha_\mathrm{d}$ $\left(a_0^3\right)$} & \multicolumn{1}{c}{$\alpha_\mathrm{q}$ $\left(a_0^5\right)$}\\
		\hline
		This work & 5.49(11) & 18(8)  \\
		\"Opik \cite{oepik} & 5.47(5) &  \\
		Risberg \cite{risbergarkfys1956}& 5.47 & 12.8 \\
		Eissa \& \"Opik \cite{eissa1967polarization} & 5.40 &  19.01 \\
		Mahan \cite{mahan1980} & 5.6  & 20\\
		Lim \textit{et al.} \cite{lim2002} & 5.515 & \\
		\hline\hline
	\end{tabular}
	\endgroup
\end{table}

The final dipole and quadrupole polarizabilities obtained by the iterative fitting procedure are listed in Table~\ref{poldq}, where they are compared to previous experimental and theoretical results, with which they are in agreement. The uncertainties of the dipole and quadrupole polarizabilities are determined by considering only the two states $39f$ and $34g$. Using Gaussian error propagation, the dependence of the intercept and the gradient on the model parameters was determined. The uncertainty of the polarization energy was chosen to be as large as the combined penetration and exchange-energy corrections. The uncertainty of the nonadiabatic expansion coefficients was chosen to be 10\,\% of the deviation from the adiabatic case ($y_0^i=1$ and $y_2^i=0$). This error analysis shows that the uncertainty of the polarizabilities is dominated by the uncertainties of the different energy corrections in Eq.~\eqref{eq:QDcontributions} and not by the accuracy of the experimentally determined quantum defects.

The inferred polarization contribution calculated using Eq.~(\ref{polfit}) and the polarizabilities obtained from fitting are given in Table~\ref{comparisiontheoryexp}.

\section{Conclusion and Outlook}

In this work, the first ionization energy of $^{39}$K was determined to be\linebreak
\num{35009.8139710(22)}$_\mathrm{sys}$(3)$_\mathrm{stat}$\,\si{\per\centi\meter} with 300-fold improved accuracy compared to the best previous result \cite{LORENZENoptcomm1981}. The uncertainty of our experimental result is limited by a systematic error in the frequency calibration. The relative uncertainty of the ionization energy is comparable to that obtained in a similar investigation in Cs \cite{deiglmayrpra2016}. Removal of the frequency calibration error in measurement series II yielded transition frequencies with an uncertainty limited by the Doppler shift arising in the spectroscopy of the ultracold atom cloud released prior to excitation. In future measurements, a systematic first-order Doppler shift could be characterized and removed, \textit{e.g.}, by choosing different propagation axes of the excitation laser. The limits on the next-leading systematic errors (DC Stark shifts and excitation-power dependent shifts) can be lowered by reducing the UV spectral linewidth (about \SI{2}{MHz}) and improving the signal-to-noise ratio of the resonances. This requires colder samples and better laser-frequency and power stability. Ultimately, our accuracy is limited by the RF reference of the optical frequency comb to \num{1e-12}, or \SI{1}{kHz}.

For the ionization energies of heavy, multi-electron systems like K and Cs, the achievable experimental accuracy currently exceeds the accuracy of theoretical \textit{ab-initio} calculations by orders of magnitude. However, applying similar techniques to lighter systems, such as H$_2$ \cite{beyerpra2018}, provides a test bed for the development of theoretical methods \cite{Hoelsch2019,Pachulski2019,matyus2018non,tung2012very} and might allow for improved determinations of fundamental constants~\cite{sprecherMeasuringIonisationDissociation2011,Hoelsch2019}.

In addition to the ionization energy, we obtained energy dependent quantum defects for the $s$, $p$, $d$, $f$ and $g$ series in $^{39}$K. This set of quantum defects in $^{39}$K improves previous work  \cite{LORENZENoptcomm1981,Lorenzenphysscrip1983,risbergarkfys1956} and can be used, \textit{e.g.}, in the calculation of Rydberg-Rydberg interaction potentials \cite{deiglmayrLongrangeInteractionsRydberg2016,weberCalculationRydbergInteraction2017} and of the energy-level structure of long-range Rydberg molecules \cite{greeneCreationPolarNonpolar2000,marcassaChapterTwoInteractions2014}. The quantum defects for $^{39}$K should, in good approximation, also describe the Rydberg series of the other naturally occurring isotopes, $^{40}$K and $^{41}$K for $n \gtrsim 30$ \cite{goyQuantumDefectsSpecificisotopicshift1986}.

The observation of the nonpenetrating high-$l$ $nf$ and $ng$ Rydberg series allowed us to determine the static dipole and quadrupole polarizabilities of the $^{39}$K$^+$ ion by employing the methodology of \citet{sansonetti1981polarization}. The determined polarizabilities strongly depend on the theoretical values of the different corrections, which ought to be calculated at higher accuracies in future work.

Using radio-frequency spectroscopy, the hyperfine splitting of the 4$s_{1/2}$ ground state of $^{39}$K was determined with a relative acuracy of \num{1e-8} (\SI{5}{Hz}) using Ramsey-type interferometry. Our value is in disagreement with the previous most precise results \cite{beckmann74,chan70}.

\begin{acknowledgments}
This work is supported financially by the Swiss National Science Foundation - National Centre of Competence in Research (NCCR) QSIT-Quantum Science \& Technology under Project No. 200020-159848, the ETH Research Grant ETH-22 15-1, the Swiss National Science Foundation (Grant 200020-172620) and the European Research Council through an advanced grant under the European Union's Horizon 2020 research and innovation programme (Grant 743121).
\end{acknowledgments}
%

\end{document}